\documentclass[aps,prl,twocolumn,superscriptaddress,showpacs]{revtex4}
\usepackage{graphics}
\usepackage{amsmath}
\usepackage{bm}

\begin{document}

\title{Electron-phonon scattering in quantum point contacts}

\author{Georg Seelig}

\affiliation{D\'epartement de Physique Th\'eorique, Universit\'e de
  Gen\`eve, CH-1211 Gen\`eve 4, Switzerland}

\author{K. A. Matveev} 

\affiliation{Department of Physics, Duke University, Box 90305, Durham, NC
  27708}

\date{November 7, 2002}

\begin{abstract}
  We study the negative correction to the quantized value $2e^2/h$ of the
  conductance of a quantum point contact due to the backscattering of
  electrons by acoustic phonons.  The correction shows activated
  temperature dependence and also gives rise to a zero-bias anomaly in
  conductance.  Our results are in qualitative agreement with recent
  experiments studying the 0.7 feature in the conductance of quantum point
  contacts.
\end{abstract}


\maketitle

The quantization of conductance in units of $G_0=2e^2/h$ observed
\cite{old} in narrow one-dimensional constrictions---the quantum point
contacts (QPC)---is well understood within a simple model of
non-interacting electrons \cite{lesovik}.  In this approach the electrons
are backscattered by the effective one-dimensional potential created by
the walls of the constriction, and the conductance follows the energy
dependence of the transmission coefficient.  In the last few years a
number of experiments
\cite{thomas,thomas1,cronenwett,kristensen,reilly,thomas2} studied the
deviations of the conductance from this picture that appear as a
shoulder-like feature at a conductance near $0.7\, G_0$.  Several
experiments have demonstrated that when a strong in-plane magnetic field
is applied, this 0.7 anomaly evolves into the spin-polarized conductance
plateau at $0.5\,G_0$ \cite{thomas,thomas1,cronenwett}.  This observation
led to the proposal that the anomaly originated from a spin-dependent
mechanism \cite{thomas}.  Subsequent theoretical attempts to understand
the 0.7 anomaly mostly followed this conjecture
\cite{wang,reimann,spivak,bruus,flambaum,rejec,hirose,spinon,meir}.

Here we concentrate on another feature of the 0.7 anomaly, namely its
strong temperature dependence.  The experiments
\cite{thomas,cronenwett,kristensen} showed that the anomalous shoulder is
weak at the lowest available temperatures, but grows when the temperature
is increased.  A detailed study \cite{kristensen} revealed that the
correction to the conductance follows an Arrhenius law
\begin{equation}
  \label{eq:arrhenius}
  \delta G\propto e^{-T_A/T}
\end{equation}
with the activation temperature $T_A$ of the order of one Kelvin.  The
activated temperature dependence implies that a certain backscattering
mechanism turns on at temperature $T\approx T_A$ and leads to partial
suppression of the conductance.  The phenomenological proposals for this
mechanism include scattering off a plasmon \cite{plasmon} or a spin wave
\cite{spinon} localized in the QPC, a low-lying spin-split subband
\cite{bruus}, as well as spin-flip scattering by a Kondo impurity
\cite{cronenwett,meir}.

In this paper we explore another backscattering mechanism, which results
in the activated temperature dependence (\ref{eq:arrhenius}), namely the
scattering of electrons by acoustic phonons.  This mechanism was discussed
in the cases of scattering by surface acoustic waves \cite{Galperin} and
by equilibrium phonons at relatively high temperatures \cite{Gurevich}.
At low temperatures the effect of electron-phonon scattering on transport
in quantum wires is strongly suppressed.  Indeed, in order to backscatter
an electron at the Fermi level the phonon must have a wavevector $q=2k_F$
in the direction along the wire.  Thus the minimum energy of such a phonon
is
\begin{equation}
  \label{eq:T_A}
  T_A = 2\hbar s k_F = \sqrt{E_s E_F}
\end{equation}
where $s$ is the speed of sound, $E_F$ is the Fermi energy of electrons in
the wire, $E_s=8ms^2$, and $m$ is the effective mass of electrons.  At
temperatures $T\ll T_A$ the occupation numbers of such phonons and their
contribution to the conductance are exponentially suppressed,
Eq.~(\ref{eq:arrhenius}).  For typical GaAs quantum wires one can estimate
$E_s\approx 0.3$K and $E_F\sim 100$K, resulting in $T_A\sim 5$K.  Thus the
electron-phonon scattering is negligible in typical low-temperature
experiments.  On the other hand, the electron density in a QPC tuned into
the vicinity of the first conductance step is very low.  Estimating the
Fermi energy at the center of a QPC to be about 2K, one obtains a value of
$T_A\approx0.8$K, in reasonable agreement with experiment
\cite{kristensen}.

To study transport in a QPC near a conductance step as a function of the
gate voltage, one has to account for the effect of the confining
potential.  We follow the idea of Refs.~\onlinecite{lesovik,buttiker} and
model the QPC by a one-dimensional electron gas in an external potential
approximated as $U(x)=-\frac{1}{2}m\Omega^2x^2$.  This approximation is
valid in the vicinity of the conductance step, provided that the potential
is sufficiently smooth.  The transmission coefficient of such a barrier is
well known \cite{Kemble},
\begin{equation}\label{transmission}
T_0(E)=\frac{1}{1+e^{-2\pi E/\hbar\Omega}}.
\end{equation}
In this paper we concentrate on the case when the Fermi energy is well
above the top of the barrier, $E_F\gg \hbar\Omega/2\pi$.  In this regime
the transmission coefficient (\ref{transmission}) at the Fermi level
equals unity up to an exponentially small correction, and even a weak
backscattering by acoustic phonons gives a significant correction
(\ref{eq:arrhenius}) to the quantized conductance $G_0$.

The amplitude of electron backscattering by a phonon with wavevector $q_x$
is proportional to the matrix element 
\begin{equation}
  \label{eq:matrix_element}
  I(q_x) = \langle \psi_R|e^{iq_x x}|\psi_L\rangle,
\end{equation}
where $\psi_{R,L}(x)$ are the wavefunctions of the right- and left-moving
electrons.  Due to the fact that the speed of sound is small, the typical
phonon energy $\hbar s q$ is much smaller than $\hbar \Omega$.  Therefore,
in evaluating the matrix element (\ref{eq:matrix_element}), the right- and
left-moving electrons can be assumed to have the same energy $E_F$.  At
$E_F\gg\hbar\Omega$ the calculation of the matrix element
(\ref{eq:matrix_element}) can be further simplified by using semiclassical
expressions for the wavefunctions:
\begin{equation}
  \label{wavefunction}
  \psi_{R,L}(x)=\sqrt\frac{m}{2\pi \hbar p_F(x)}\,
                  \exp\left[\pm\frac{i}{\hbar}\int_0^x dx'p_F(x')\right].
\end{equation}
Here the quasiclassical Fermi momentum is defined as
$p_F(x)=\sqrt{2m[E-U(x)]}$.

We begin by studying the regime of small temperature and bias, $T,eV\ll
T_A$, when the dominant contribution to $\delta G$ is due to
long-wavelength phonons, $q_x\ll 2k_F$.  (Here $k_F=\sqrt{2mE_F}/\hbar$ is
the electron wavevector at the center of the QPC.)  In this regime the
integral (\ref{eq:matrix_element}) is easily evaluated by the saddle point
method, and we find
\begin{eqnarray}
  \label{qsmall}
  I(q_x)&=&\left(\frac{k_F^2}
            {2\pi\hbar\Omega E_Fq_x\sqrt{4k_F^2-q_x^2}}\right)^{1/2}
            e^{-F_1(q_x/2k_F)},\ \ \\
  F_1(u)&=&\frac{2E_F}{\hbar\Omega}\left(
              \arccos u-u\sqrt{1-u^2}\right).
             \nonumber
\end{eqnarray}
Since $E_F$ is assumed to be large compared to $\hbar\Omega$, this result
is valid even when $q_x$ approaches $2k_F$, as long as $F_1\gg1$.

At zero temperature only the processes of emission of phonons are
allowed.  The energy of the phonon cannot exceed applied voltage, and thus
the maximum possible $q_x$ in such a process is $eV/\hbar s$.  Therefore,
at $eV<T_A$ the correction to the differential conductance is
exponentially small as $\delta G\propto \exp[-2F_1(eV/T_A)]$.

At non-zero temperature, both emission and absorption of phonons are
possible.  In the most interesting limit of zero bias, both processes are
exponentially suppressed at $T\ll \hbar s q$.  Thus the total
backscattering rate is small as $\exp[-2F_1(q_x/2k_F) -\hbar s q_x/T]$.
One can easily see that this rate assumes its maximum value at
$q_x=2k_F\sqrt{1-(\hbar\Omega T_A/8E_FT)^2}$.  Thus,
the temperature dependence of the correction to the quantized conductance
has the following exponential behavior:
\begin{equation}
  \label{eq:temperature_dependence}
  \delta G =-G^* \exp\left[-\frac{T_A}{2T}\left(
                             \frac{\arcsin v}{v}+\sqrt{1-v^2}
                   \right)\right],
\end{equation}
where $v=\hbar\Omega T_A/8E_FT$.
In a broad range of temperatures $\frac{\hbar\Omega}{E_F} T_A\ll T\ll T_A$ this
correction shows activated behavior (\ref{eq:arrhenius}).  The deviation
from the Arrhenius law (\ref{eq:arrhenius}) at low $T$ is due to the
presence of the potential barrier $-\frac12 m\Omega^2x^2$ for the
one-dimensional electrons in the QPC.

In order to find the preexponential factor $G^*$ in
Eq.~(\ref{eq:temperature_dependence}) and to evaluate $\delta G$ in the
regime when either temperature or voltage exceed $T_A$, one
has to perform a more formal calculation of the electron backscattering
rate.  Using the golden rule approach, one can find the following
expression for the scattering rate of a right-moving electron of energy
$E$ to all the available left-moving states:
\begin{eqnarray}\label{rate}
\tau_{R}^{-1} (E)&=&2\pi\sum_{\lambda}\int\frac{d\bm q}{(2\pi)^3}
                     \frac{|M_\lambda(\bm{q})|^2}
                          {2\rho\omega_{\bm{q}\lambda}}
                     \int dE' |I(q_x)|^2\nonumber\\
                   &&\times[1-f_L(E')]\{N(\omega_{\bm{q}\lambda})
                            \delta(E-E'+ \hbar\omega_{\bm{q}\lambda})
                        \nonumber\\
                   &&+[N(\omega_{\bm{q}\lambda})+1]
                            \delta(E-E'-\hbar\omega_{\bm{q}\lambda})\},
\end{eqnarray}
Here $\lambda$ labels the three possible polarizations of the acoustic
phonons, $\omega_{\bm{q}\lambda} \propto q$ is the phonon frequency,
$\rho$ is the mass density of the material, $f_L(E')$ is the Fermi
function of the left-moving electrons in the contact,
$N(\omega_{\bm{q}\lambda})$ is the equilibrium occupation number of a
phonon of wavevector $\bm{q}$ and polarization $\lambda$.

The exact form of the matrix element $M_\lambda(\bm{q})$ depends on the
nature of the electron-phonon coupling.  At relatively high temperatures
\cite{Gurevich} the deformation potential coupling dominates, and
$M_\lambda(\bm{q})\propto q$.  However, at low temperatures in GaAs the
leading contribution is due to piezoelectric coupling, for which
$M_\lambda(\bm{q})$ depends on the direction of $\bm{q}$, but not on its
length $q$.

The backscattering of electrons by phonons reduces the total current
carried by the right-moving electrons of energy $E$ in the contact.  One
can therefore find the correction to the total current $I=G_0V$ by
integrating the backscattering rate (\ref{rate}) with the occupation
numbers of the respective states:
\begin{equation}
 \label{eq:current_corection}
 \delta I=-2e\int^\infty_{-\infty}[\tau_{R}^{-1}(E)f_R(E)
                                   -\tau_{L}^{-1}(E)f_L(E)]dE.
\end{equation}
Here the factor of 2 accounts for the electron spins; the expression for
the scattering rate $\tau_L^{-1}$ of left-moving electrons is obtained
from Eq.~(\ref{rate}) by replacing the subscripts $R\leftrightarrow L$.

The correction $\delta I$ depends on the voltage $V$ across the contact
through the difference of the chemical potentials $\mu_R-\mu_L=eV$
entering the Fermi functions $f_R(E)$ and $f_L(E)$.  Thus, the correction
to the conductance $G_0$ of the contact can be found as $\delta G=d\delta
I/dV$.  The resulting expression for $\delta G$ is simplified greatly if
one makes the following approximations.  First, we again neglect the
dependence of the matrix element (\ref{eq:matrix_element}) on the energies
$E$ and $E'$, and assume $E=E'=E_F$.  Second, following the standard
procedure \cite{ridley}, we replace $|M_\lambda(\bm{q})|^2$ and the sound
velocities with their values $|M_\lambda|^2$ and $s_\lambda$ averaged over
the directions of $\bm{q}$.  Then the integrals with respect to the
energies $E$ and $E'$ as well as the transverse components of $\bm{q}$
can be done analytically, and we find
\begin{equation}
  \label{result0}
  \delta G=-G_0T\sum_\lambda\frac{|M_\lambda|^2}{2\rho s_\lambda^2}
                \int^\infty_{-\infty}dq_x|I(q_x)|^2K_\lambda(q_x).
\end{equation}
Here the function $K_\lambda(q_x)$ is given by 
\begin{eqnarray}
K_{\lambda}(q_x)&=& 2\ln\frac{1}{1-e^{-\hbar  s_{\lambda}|q_x|/T}}
                     -\frac{\hbar s_{\lambda}|q_x|}{T}
                    \nonumber\\
&&+\frac{eV-\hbar s_\lambda|q_x|}{2T}
    \coth \frac{eV-\hbar s_{\lambda} |q_x|}{2T}
    \nonumber\\
&&+\frac{eV+\hbar s_\lambda|q_x|}{2T}
    \coth \frac{eV+\hbar s_{\lambda} |q_x|}{2T}.
\end{eqnarray}
Unlike our previous results, the expression (\ref{result0}) for the
correction $\delta G$ is not limited to the regime $T, eV\ll T_A$.  In
addition, when the temperature and voltage are small compared to $T_A$,
Eq.~(\ref{result0}) enables one to find the preexponential factors, such
as $G^*$ in Eq.~(\ref{eq:temperature_dependence}).

To find $G^*$, we first notice that at $T\ll T_A$ the longitudinal phonon
mode can be ignored.  Indeed, the sound velocity $s_t$ of the transverse
modes is lower than that of the longitudinal mode, $s_t<s_l$.  Accordingly,
the activation temperature (\ref{eq:T_A}) is lower for the transverse
modes, i.e., the longitudinal mode gives a negligible contribution to
$\delta G$ at low temperatures.  We will therefore assume $s=s_t$ in the
definition (\ref{eq:T_A}) of the activation temperature.

At zero bias and $T\ll \hbar s_t |q_x|$, we find $K(q_x)=(2\hbar s_t
|q_x|/T)e^{-\hbar s_t |q_x|/T}$.  The integral in Eq.~(\ref{result0}) can
then be evaluated by the saddle-point method.  As a result we reproduce
the exponential temperature dependence (\ref{eq:temperature_dependence})
with the prefactor
\begin{equation}
  \label{eq:prefactor}
  G^*=\gamma G_0 
          \sqrt{\frac{2T}{\pi T_A}}\frac{1}{\sqrt[4]{1-v^2}},
  \quad \gamma=\frac{2|M_t|^2}{\rho s_t}\frac{m}{\hbar^2\Omega}.
\end{equation}
The dimensionless parameter $\gamma$ determines the magnitude of the phonon
backscattering effect on conductance at $T=T_A$.  The numerical value of
$\gamma$ can be estimated from the data available in the literature
\cite{ridley}.  We write the coupling parameter for the transverse phonons
as $|M_t|^2=\frac{8}{35}(e e_{14}/\epsilon)^2$, where for GaAs the
permittivity $\epsilon=13.2\, \epsilon_0$, and $e_{14}=0.16$ C/m$^2$.
Substituting $\rho=5.36\,{\rm g/cm^3}$, $s=s_t=3\times 10^{3}\, {\rm
  m/s}$, and $m=0.067\, m_e$, we find $\gamma = 0.0052\, {\rm
  meV}/\hbar\Omega$.

We now turn to the evaluation of the correction (\ref{result0}) to the
conductance of the QPC in the regime when the temperature and/or bias are
large compared to $T_A=2\hbar s_tk_F$.  In this case the typical
wavevector $q$ of the phonons emitted or absorbed by electrons is large,
$q\gg k_F$.  To find $\delta G$ we notice that the matrix element $I(q_x)$
has a peak near $q_x=2k_F\ll q$.  Thus, the electron backscattering in
this regime is dominated by phonons propagating in the direction normal to
the channel.  One can then substitute the asymptotic expression 
for $K_\lambda$ at
$q_x\to0$ into Eq.~(\ref{result0}) and find $\delta G$ in the form
\begin{equation}
  \label{result2}
  \delta G=-\tilde\gamma G_0\left(\frac{T}{T_A}\ln\frac{T}{T_A}
                                 +\frac{eV}{2T_A}\coth\frac{eV}{2T}\right).
\end{equation}
Here the dimensionless parameter $\tilde\gamma$ is defined as
\begin{displaymath}
  \tilde\gamma = \sum_\lambda \frac{|M_\lambda|^2 s_t}{\rho s_\lambda^2}
                             \frac{m}{\hbar^2\Omega}.
\end{displaymath}
Due to the contribution of the longitudinal phonon mode,
$\tilde\gamma>\gamma$.  To estimate $\tilde\gamma$ we write the average
matrix element as $|M_l|^2=\frac{12}{35}(e e_{14}/\epsilon)^2$,
Ref.~\cite{ridley}. Then using the value $s_l=5.12\times10^{3}\, {\rm
  m/s}$ of the velocity of longitudinal sound in GaAs, we find
$\tilde\gamma = 0.0065\, {\rm meV}/\hbar\Omega$.

It is interesting to note that the negative correction (\ref{result2}) to
the quantized conductance $G_0$ grows not only with temperature, but also
with bias.  When $V$ is increased, more left-going states become available
for the right-moving electrons to scatter into, and the conductance
decreases.  Thus, the electron-phonon scattering gives rise to a zero bias
anomaly similar to the one observed in experiments
\cite{cronenwett,kristensen}.  The linear shape of the zero-bias peak at
$eV\gg T$ is consistent with the one observed in
Ref.~\onlinecite{cronenwett}.  The height of the peak is determined by the
limits of applicability of Eq.~(\ref{result2}) at high bias.  The most
important limitation of our derivation was the assumption that the
electrons are purely one-dimensional.  At sufficiently high bias the
typical wavevector $q\sim eV/\hbar s$ of the phonons becomes comparable to
the width $w$ of the one-dimensional channel.  Since the backscattering is
mostly due to the phonons propagating in the transverse direction, their
coupling to the electrons at $q>1/w$ becomes weak, and the linear
dependence $\delta G(V)$ given by Eq.~(\ref{result2}) saturates.  This
saturation occurs at $eV\sim T_A\sqrt{\Delta/E_F}$, where
$\Delta\sim\hbar^2/mw^2$ is the subband spacing in the QPC.  Thus the
height of the zero-bias peak in conductance is expected to be of the order
$\tilde\gamma G_0 \sqrt{\Delta/E_F}$.

The zero-bias peak observed in experiment \cite{cronenwett} had a height
of about $0.15\, G_0$.  To compare this result with our estimate
$\tilde\gamma G_0 \sqrt{\Delta/E_F}$, we use the device parameters
$\hbar\Omega\approx 0.3\,{\rm meV}$ and $\Delta\approx0.9\, {\rm meV}$
reported for similar samples \cite{taboryski}.  To estimate $E_F$ we
assume the transmission coefficient $T_0(E_F)\approx0.9$ and from
Eq.~(\ref{transmission}) find $E_F\approx \frac13\hbar\Omega$.  This
results in the peak height $\tilde\gamma G_0 \sqrt{\Delta/E_F}\approx
0.07\,G_0$, which is reasonably close to the experimentally observed value
$0.15\, G_0$.   

Unlike the bias dependence of $\delta G$, the temperature dependence
$\delta G = -\tilde\gamma G_0 (T/T_A)\ln(T/T_A)$ obtained from
Eq.~(\ref{result2}) at $V=0$ does not saturate at $T\sim
T_A\sqrt{\Delta/E_F}$.  The suppression of coupling to phonons with
$q>1/w$ does cut off the factor $\ln(T/T_A)$.  However, the main linear in
$T$ dependence of $\delta G$ originates from the phonon occupation numbers
$N(\omega_q) = T/\hbar\omega_q$ at $T\gg\hbar\omega_q$, and remains even
at $T \gg T_A \sqrt{\Delta/E_F}$ \footnote{The correction (\ref{result2})
  will eventually saturate at $|\delta G|\sim G_0$ due to the higher-order
  terms in the electron-phonon coupling constant.}.  The experiment
\cite{cronenwett} does show a stronger suppression of conductance $\delta
G\sim -0.3\, G_0$ at high temperature than $\delta G\sim -0.15\, G_0$ at
high bias.  It is also worth noting that a device with a higher value of
$\hbar \Omega=2.6\,{\rm meV}$ and, consequently, lower $\tilde\gamma$
shows a weaker temperature dependence of $\delta G$,
Ref.~\onlinecite{kristensen1}.

A quantitative comparison of the effect of electron-phonon backscattering
with experiments should account for the Coulomb interactions between the
electrons.  In two- and three-dimensional systems the main effect of the
interactions upon the electron-phonon scattering is due to the screening,
which leads to suppression of coupling at low energies \cite{ridley}.  On
the contrary, in quantum point contacts the electron-phonon scattering
should be \emph{enhanced} by the Coulomb interactions between electrons.
Indeed, it is well known \cite{kane} that the backscattering probability
for an electron in an interacting one-dimensional system is enhanced by a
factor $(D/T)^{1-g_\rho/2}$.  Here $D\sim E_F$ is the bandwidth, and
$g_\rho$ is the parameter describing the strength of the interactions in
the Luttinger liquid.  The one-dimensional electron gas at the center of a
quantum point contact at gate voltage corresponding to the first quantized
step of conductance is extremely dilute.  Consequently, the Coulomb
interactions of electrons are very strong, and $g_\rho\ll1$.  Thus, we
expect the correction $\delta G$ to be enhanced by a large factor of order
$E_F/T_A$ due to the Coulomb interactions.

The phonon-induced backscattering effect discussed in this paper is not
limited to the first conductance step.  Although most experiments observe
the anomalous shoulder in the conductance at the first step, several
observations of similar behavior at the second
\cite{thomas1,reilly,kristensen1} and even higher steps \cite{kristensen1}
have been reported.

By applying an in-plane magnetic field $B$ one can polarize the electron
spins and observe a conductance step of height $0.5 G_0$.  The
phonon-induced backscattering should then result in a negative correction
to conductance similar to the 0.7 anomaly at $B=0$.  However, the
experiments \cite{thomas,thomas1,cronenwett,thesis} do not show an
anomalous plateau at $0.7\times(0.5 G_0)$.  The likely reason for the
apparent absence of the phonon backscattering effects is that at
temperatures $T\gtrsim T_A$ the electrons in the channel are no longer
completely spin-polarized.  Indeed, in the experiment \cite{thesis} the
spin-split conductance plateau at $G=0.5G_0$ \emph{rises} to values about
$0.6 G_0$ when the temperature is increased from 80mK to 1.3K, indicating
that the second spin-split subband contributes to the conductance.  This
conclusion is supported by the estimate of the spin-splitting
$g\mu_BB\approx 3$\,K in a typical field $B=10$T.  Thus, at temperature of
order $T_A\sim 1K$ the second spin-split subband gives a significant
positive contribution to the conductance that compensates for the decrease
in conductance due to the phonons.  To observe the phonon-induced
backscattering features in conductance, magnetic fields significantly
higher than 10T are required.

In conclusion, we have studied the effect of backscattering of electrons
in quantum point contacts by acoustic phonons.  We found a significant
negative correction to the quantized conductance.  The correction grows
exponentially as a function of temperature or voltage at $T,eV\ll T_A$.
Above the activation temperature $T_A$, the correction grows roughly
linearly with $T$ and $V$, Eq.~(\ref{result2}).  Our results are
consistent with the experimentally observed features of the conductance
near $0.7(2e^2/h)$.
  
\begin{acknowledgments}
  We would like to acknowledge helpful discussions with A.~V. Andreev,
  L.~S. Levitov, V.~I. Fal'ko, C.~M. Marcus, R. de Picciotto, and the
  hospitality of the Bell Laboratories where most of this work was carried
  out.  GS acknowledges the support of the Swiss National Science
  Foundation, the Swiss Study Foundation and the Dr. Max Husmann-Stiftung.
  KAM acknowledges the support of the Sloan Foundation and NSF Grant
  DMR-0214149.
\end{acknowledgments}

\end{document}